# $d^0$ Ferromagnetism in Mg-doped Rutile $TiO_2$ Nanoparticles

L. Chouhan[1], G. Bouzerar[2] and S. K. Srivastava[1*]

[1]Department of Physics, Central Institute of Technology Kokrajhar, Kokrajhar-783370, India
[2]CNRS et Université Claude Bernard Lyon 1, F-69622, Lyon, France

[*]**Corresponding Author E-mail:** sk.srivastava@cit.ac.in

## Abstract

In a quest of enriching the area of $d^0$ magnetism in oxide materials, we have undertaken to study Mg-doped $TiO_2$ compounds. The $Ti_{1-x}Mg_xO_2$ (x=0, 0.02, 0.04 and 0.06) nanoparticles were prepared by solid-state reaction route. The X-ray diffractions (XRD) patterns of these samples indicate single phase of tetragonal rutile-structure of $TiO_2$. The refinement of the XRD patterns reveals no change in the crystallographic lattice parameters in comparison to pure $TiO_2$ upon Mg doping and it indicates that $Mg^{2+}$ ions do not enter core grains and form core/shell structure. SEM observations reveal the uniform morphology with nanometric grains in the range of 150-200 nm. The measurement of magnetic properties of these compounds indicates that pure $TiO_2$ and $Ti_{0.98}Mg_{0.02}$ compounds exhibit paramagnetic behavior and $Ti_{0.96}Mg_{0.04}$ compound exhibits ferromagnetic (FM) phase superimposed with the dominating paramagnetic phase. However, $Ti_{0.94}Mg_{0.06}$ compound exhibits ferromagnetic to paramagnetic transition with FM transition temperature of 180.2 K. The measurements of zero field and field cooled magnetization data indicate low temperature magnetic irreversibility for x=0.06 sample and it was attributed to the competing AFM (core) and the FM (shell) interactions. The measurement of hysteresis curves at various temperatures indicates domain wall pinning and an exchange-bias behavior.

**Key Words:** Rutile $TiO_2$; Mg-doping; $d^0$ Ferromagnetism; Core-Shell Magnetism; Domain wall pinning; Exchange Bias.

# 1. Introduction:

Since many years, researchers are putting their persistent effort to explore new materials that could transport a spin-polarized electric current and can be integrated in spintronics devices, such as giant magneto-resistance sensors, magneto-resistive random-access memories and storage media [1, 2]. Among the various explored materials for spintronics devices, transition metal (TM) doped semiconducting oxide materials were studied extensively [3-7] in the past. But it has not yet resulted in reproducible and homogeneous magnetic materials and there is a debate whether the observed magnetism is intrinsic or due to the magnetic ion cluster [8]. In order to get a clean material exhibiting room temperature ferromagnetism and that without any transition metal atoms, several other types of materials were studied and explored. Recently, non-magnetic element doped induced magnetism in oxide materials or so called, $d^0$ ferromagnetism was believed to provide an alternative pathway to TM-induced ferromagnetism [9]. In a model proposed by Bouzerar *et al.*, it was predicted that ferromagnetism (FM) with high Curie temperature are feasible by substitution of non-magnetic elements such as, Li, Na, K, etc. in oxide matrix [10]. In-fact, many *ab-initio* studies have predicted ferromagnetism with high Curie Temperature ($\theta_C$) in several non-magnetic elements doped oxides, such as K [11], Ag [12], Mg [13] doped $SnO_2$; Li-doped anatase $TiO_2$ [14], K-doped rutile $TiO_2$ [15, 16], V-doped $TiO_2$ [17], Mg-doped $TiO_2$ [18]; K [15], Cu [19], V [20] and Ag [21-23] doped zirconia; non-magnetic element-doped ZnO [24-28]. Experimentally, $d^0$ magnetism were observed in several non-magnetic elements doped oxides such as; Cu doped $TiO_2$ [29, 30], C-doped $TiO_2$ [31], K-doped $SnO_2$ [32], Li-doped $SnO_2$ [33], K-doped $TiO_2$ [34], Na-doped $SnO_2$ [35] and Mg-doped $ZrO_2$ [36], such as Cu-doped ZnO [28], alkali metal doped ZnO [37] Li doped-ZnO [38-43] and Nd-doped ZnO nanowires [44].

Among many potential oxide materials, titanium dioxide ($TiO_2$) has attracted lot of attention, as it possesses good chemical stability, excellent optical, catalytic, thermal and electronic properties. It is used in various scientific & technological applications like photocatalysis, sensor, solar cell material, antifogging devices, self-cleaning coating and so on [45-48]. $TiO_2$ is found to exist in three crystalline forms: anatase, rutile, and brookite. Out of the three crystalline forms rutile is the most abundant natural form of $TiO_2$, which has tetragonal structure (space group $P4_2/mnm$) [49-50]. Several reports indicate existence of RTFM in undoped $TiO_2$ films, single crystals, bulk powder and nanoribbons [50-54]. The coexistence of Ti vacancies ($V_{Ti}$) and O vacancies ($V_o$) is considered as one possible origin of the ferromagnetic ordering in undoped rutile TiO$_2$ [55]. Moreover, many theoretical studied [14-18] have predicted and experimental studies [34-36] have demonstrated d$^0$ ferromagnetism in non-magnetic element doped TiO$_2$. In a recent theoretical investigation (using the LDA+U method) on magnetic and optical properties of Mg-doped anatase TiO$_2$ [18], it was predicted that these materials can induce moments about 2 $\mu_B$, which mainly localize on the nearest apical oxygen atoms. Moreover, it indicates that the magnetic moments prefer to the ferromagnetic coupling with the antiferromagnetic state [18]. In pursuit of augmenting the research area of $d^0$ ferromagnetism and motivated by this theoretical calculation [18], we have endeavoured an experimental study on the crystal structure and magnetic properties of Mg-doped TiO$_2$ compounds. We have chosen to prepare materials in bulk form at equilibrium conditions to diminish the uncertainties in fabrications and any inaccuracies in characterization.

## 2. Experimental Details

The polycrystalline samples of Ti$_{1-x}$Mg$_x$O$_2$ (x=0, 0.02, 0.04 and 0.06) were prepared by solid-state reaction route. We used high-purity TiO$_2$ and MgO as the starting materials for the synthesis of the samples. Pre-sintering of the samples was performed in powder form at various temperatures, i.e. 300$^0$C, 500$^0$C and 1000˚C for about 10 hours at each temperature. The samples were finally annealed in pallet form at 1250$^0$C for 30 hrs. The structure and phase purity were checked using high resolution x-ray powder diffraction. Micro-structural images have been carried out by a scanning electron microscope (SEM). The temperature (T) variation of magnetization (M) and, magnetization versus magnetic field (H) measurements were carried out using commercial SQUID magnetometer (Quantum Design, MPMS).

## 3. Results and Discussion

The crystal structure and phase purity of all Mg-doped TiO$_2$ compounds were checked by X-ray diffractometer and the XRD patterns of these compounds are shown in Figure 1. All the diffractions peaks could be indexed on the basis of the tetragonal rutile type-structure. It also indicates that samples are formed in single phase and there is no crystallized secondary phase. Nevertheless, no clear shift of the peaks in comparison to pure TiO$_2$ is observed. The refinement of the XRD patterns was performed with the help of the Fullprof program by using the Rietveld refinement technique [56] to estimate various crystal structural parameters. Figure 2 shows the typical Rietveld refinement of XRD patterns for Ti$_{0.98}$Mg$_{0.02}$O$_2$ compound. It is seen that the experimental XRD data matches perfectly with the Rietveld software generated XRD data. The refinement of XRD data reveals no change in the crystallographic lattice parameters in comparison to pure $TiO_2$ upon Mg doping. The lattice parameters for pure $TiO_2$ was found to be a = b = 4.5945 Å and c = 2.9586 Å, and are comparable to those reported in other work [34]. This observation strongly indicates that no solid solution is formed i.e. that $Mg^{2+}$ ions do not enter grain cores and natural core/shell structure is formed. This can be understood due to the substitution of bigger ion of $Mg^{2+}$ (0.72 Å) in 6-coordinate $Ti^{4+}$ (0.61 Å) ion. The crystallites size of pure $TiO_2$ is evaluated from XRD and it is found to be in the range of 32 nm. However, the mean diameter of crystallites size of Mg-doped sample appears to increase and it is found to be in the range of 50-70 nm. To understand the microstructure of prepared samples, we have performed observations of samples morphology using SEM. Two typical SEM images of $Ti_{0.98}Mg_{0.02}O_2$ and $Ti_{0.94}Mg_{0.06}O_2$ compounds are shown in Figure 3. SEM observations reveal the uniform morphology with nanometric grains in the range of 150-200 nm. The grain size is found to increase for higher Mg concentration of Mg: $TiO_2$ compounds.

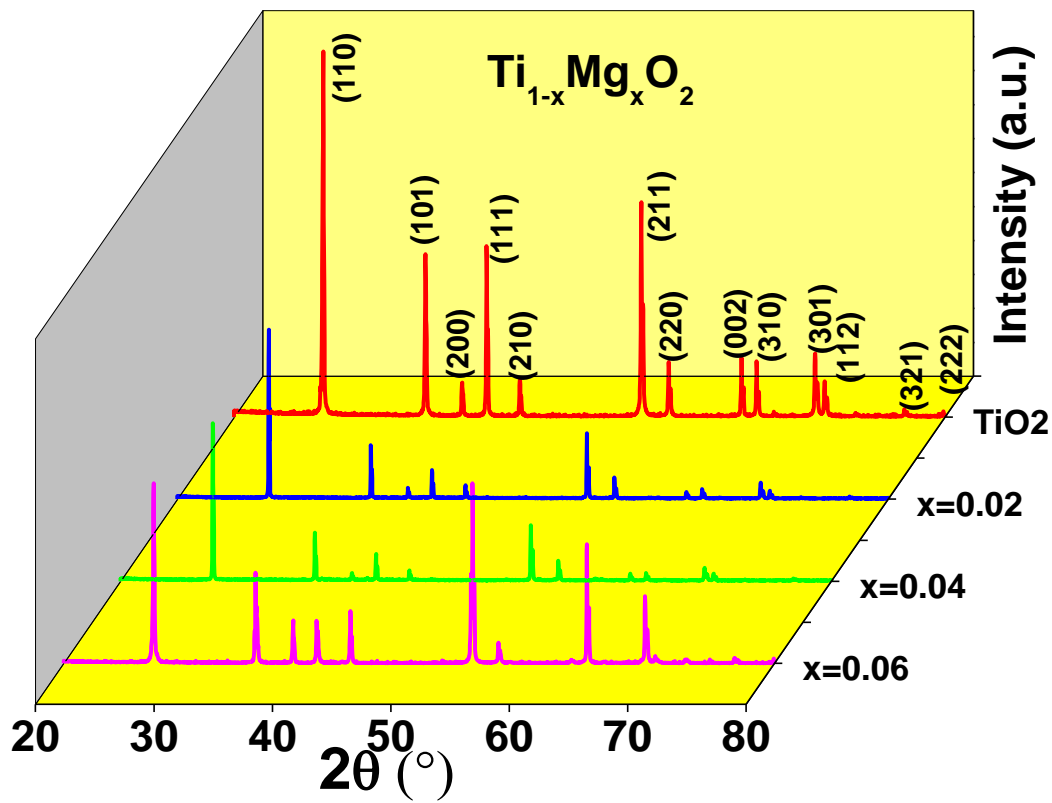

**Figure 1:** XRD patterns of $Ti_{1-x}Mg_xO_2$ (x=0, 0.02, 0.04 and 0.06) compounds.

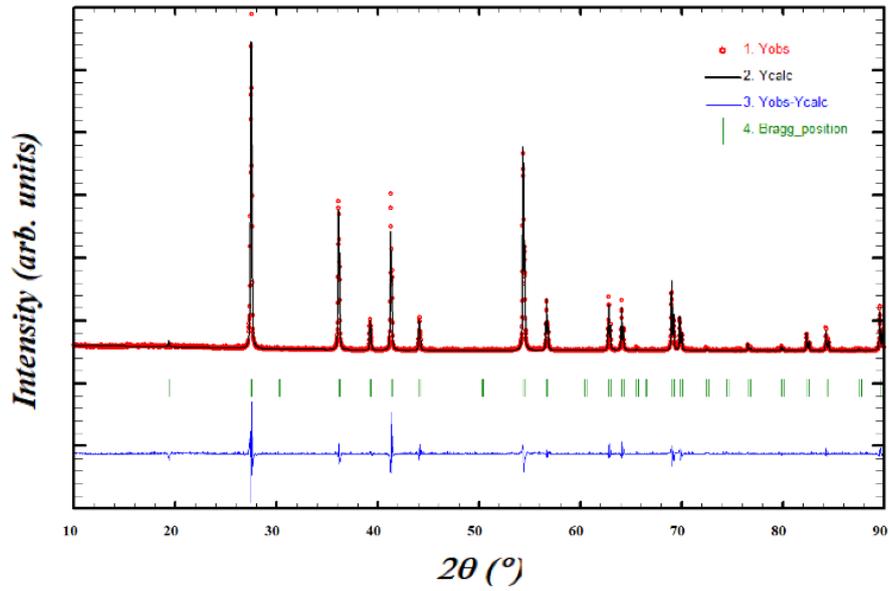

**Figure 2:** Refinement of XRD patterns for $Ti_{0.98}Mg_{0.02}O_2$ compound, obtained with the help of the Fullprof program by employing the Rietveld refinement technique.

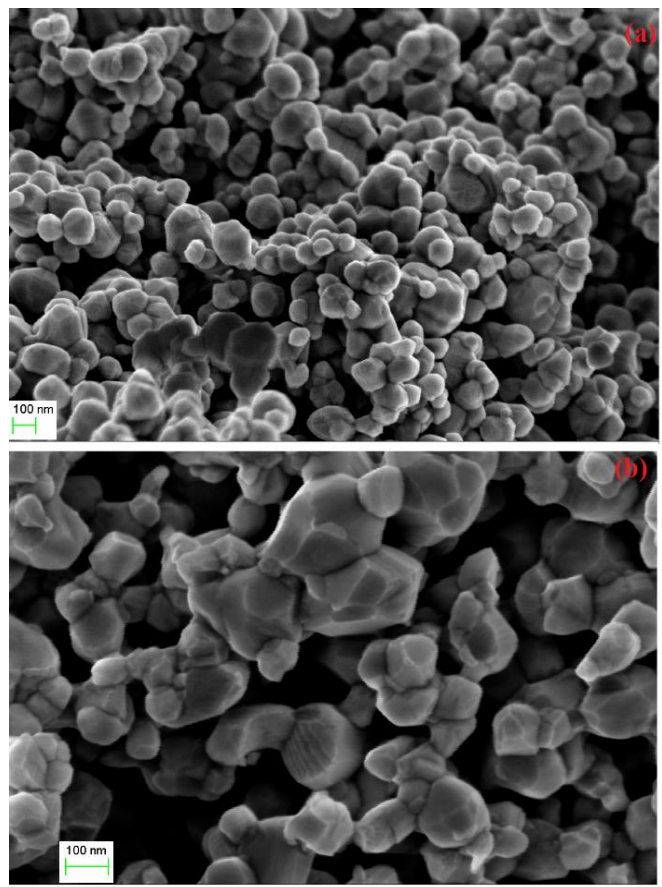

**Figure 3:** SEM images of $Ti_{0.98}Mg_{0.02}O_2$ and $Ti_{0.94}Mg_{0.06}O_2$ compounds.

To explore the magnetic properties of these samples, the zero-field cooled (ZFC) magnetization curves as a function of temperature for all Mg-doped $TiO_2$ compounds were measured using SQUID magnetometer under an applied field of 500 Oe and for a temperature range of 3-300 K. The variation of magnetization as a function of temperature i.e. M-T curve for pure $TiO_2$ and $Ti_{0.98}Mg_{0.02}O_2$ compounds show that the magnetization decreases with the increase of temperature throughout the measured temperature range i.e. an observation of paramagnetic behavior of the sample, as depicted in Figure 4(a) and (b). Thus, pure $TiO_2$ and $Ti_{0.98}Mg_{0.02}O_2$ compound is found to exhibit paramagnetic behavior. However, the measurement of M-T curve of $Ti_{0.96}Mg_{0.04}O_2$ compound indicate paramagnetic behavior along-with a low temperature ferromagnetic (FM) transition at ~65 K, indicating that weak ferromagnetic phase is superimposed with the dominating paramagnetic phase. Further, the highest Mg-doped sample i.e. $Ti_{0.94}Mg_{0.06}O_2$ compound is found to exhibit a clear ferromagnetic to paramagnetic transition. The ferromagnetic transition temperature was estimated from the lowest peak observed in |dM/dT| versus temperature plot as shown in the inset of Figure 4(d). The value is found to be 180.2 K. The broadening of the ZFC peak reveals a distribution for the crystallite size, which is consistent with our observations on the particle morphology of this system. Similar behavior was observed by Oliveira et al. [57]. To get further insight into the magnetic property, we have measured the M-T curve under field cooled (FC) condition along with ZFC condition for pure $TiO_2$, $Ti_{0.98}Mg_{0.02}O_2$ and $Ti_{0.96}Mg_{0.04}O_2$ compounds. The measurement of ZFC and FC data for pure $TiO_2$ reveals typical paramagnetic characteristic of the sample. However, the ZFC and FC M-T curves for $Ti_{0.96}Mg_{0.04}O_2$ sample coincide at low temperature and the magnitude of magnetization at FM transition temperature under FC condition has increased, indicating a ferromagnetic phase embedded in the paramagnetic phase. The measurements of ZFC and FC magnetization data for $Ti_{0.94}Mg_{0.06}O_2$ sample indicate low temperature magnetic irreversibility in ZFC and FC curves, which suggests presence of two competing magnetic phases i.e. antiferromagnetic (AFM) phase and ferromagnetic (FM) below irreversibility temperature. Such competing interactions may lead to spin-canted ferromagnetic blocks or to antiferromagnetically coupled ferromagnetic blocks. Considering all the previous results obtained, including the crystal structure parameters variation, morphology of the nanoparticles and increase in the magnitude of magnetization with Mg concentration, it is plausible to attribute the two different magnetic phases to two different regions in the nanoparticles: the core and the shell, as depicted in Figure 5. The former is an AFM nanophase, which orders progressively under reducing temperature. The latter is a surface layer with uncompensated spins and the spins are blocked below irreversibility temperature (Blocking Temperature) and, fluctuates with the increase of temperature. This core–shell model is consistent with experimental data obtained in similar kind of investigation [57-58].

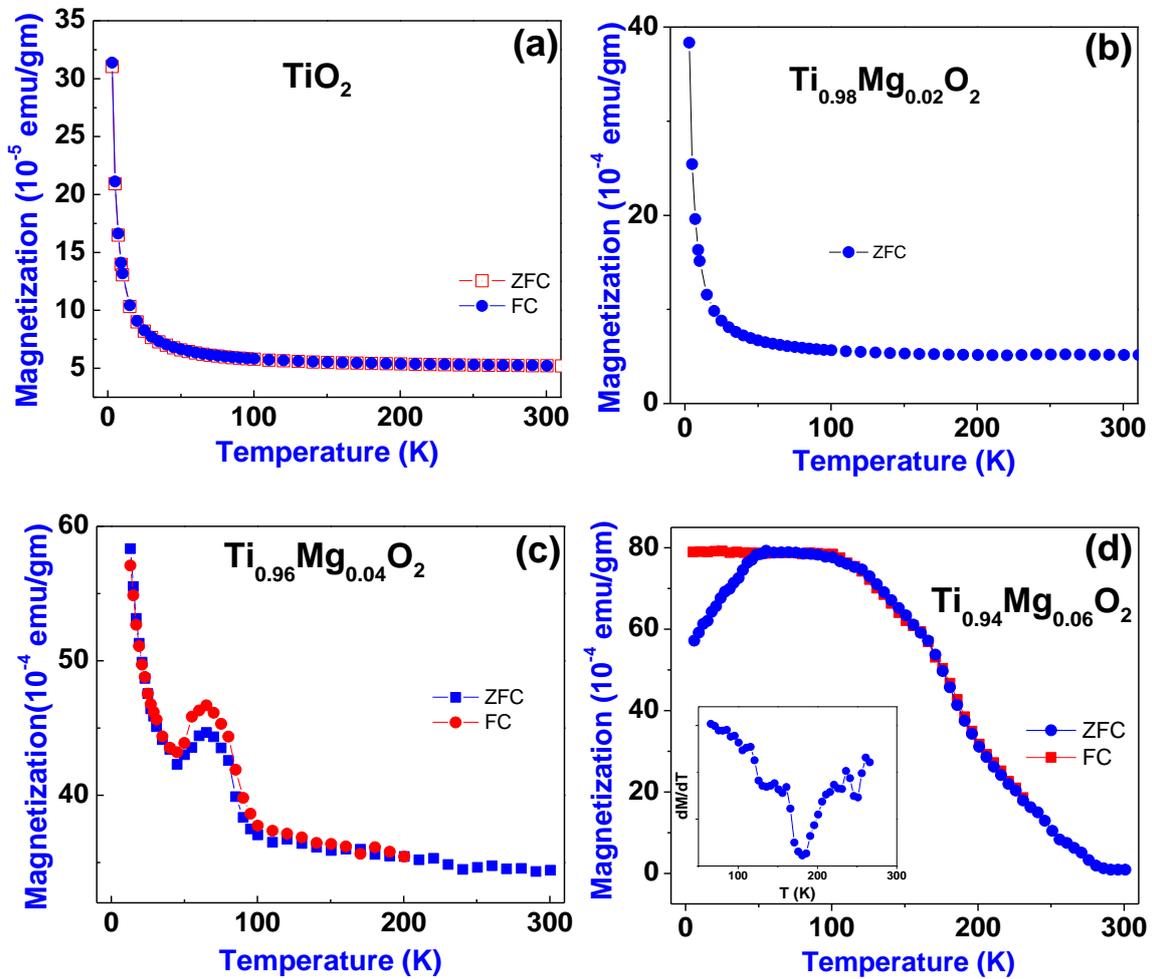

**Figure 4:** (a) Temperature variation of magnetization of Mg-doped $TiO_2$ samples measured under an applied field of 0.05 T field.

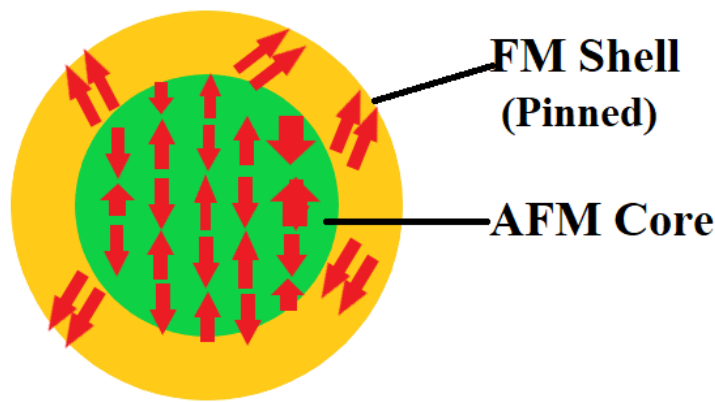

**Figure 5:** Core-shell model of $Ti_{1-x}Mg_xO_2$ compounds.

To get further understanding about the magnetic properties, we have measured field dependence of magnetization i.e. M-H data in the field range of ± 5 T at 5 K and 300 K and they are presented in Figure 6. The M-H measurements show that pure $TiO_2$ and 2% Mg doped compounds exhibit paramagnetic behavior down to 5 K, as shown in Figure 6 (a) and (b). The strong linear behaviour of M-H curve measured for $Ti_{0.96}Mg_{0.04}O_2$ at low temperature (5 K) is possibly due to the presence of weak FM phase along-with paramagnetic phase in the sample. However, the $Ti_{0.94}Mg_{0.06}O_2$ compound clearly exhibit ferromagnetic hysteresis loop at 5 K, which is not saturated up to H = 5 T and it can be interpreted as a sum of two contributions and can be represented as $M(H) = M_{FM}(H) + \chi_{AFM}H$. The first term accounts for a ferromagnetic phase component, and the second term refers to a linear AFM component of the magnetization. Such a system is expected to be a paramagnet above blocking temperature, as observed from M-H curve at 300 K. The estimation of FM component gives rise to a saturation magnetization with value of $0.05\mu_B$/f.u. as shown in the left inset of Figure 6 (b). Moreover, a vertical pinning/shifting during reversal of applied magnetic field of in the M-H curve is observed, as shown in the right inset of Figure 6 (b). This is likely due to the presence of two magnetic phases in contact i.e. core/shell structure.

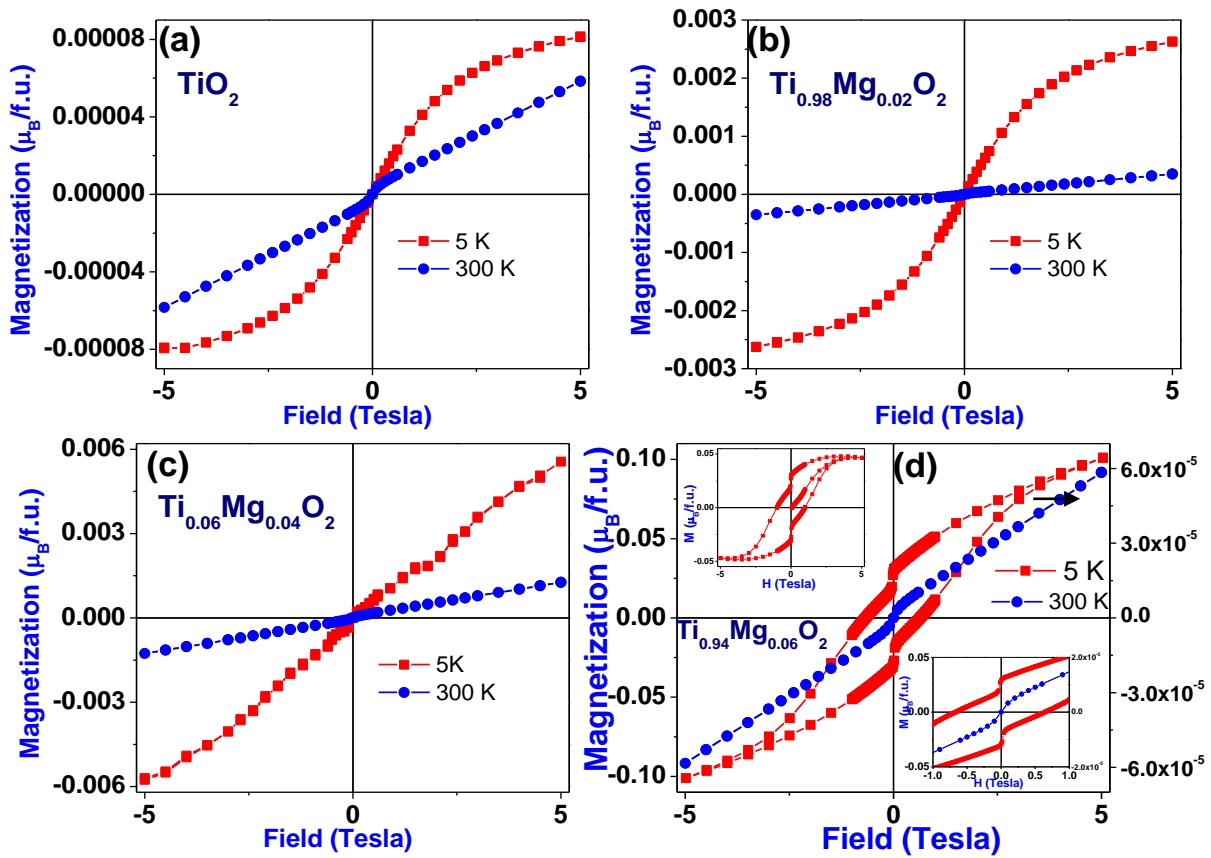

**Figure 6:** Field variation of magnetization for $Ti_{1-x}Mg_xO_2$ (x=0, 0.02, 0.04, and 0.06) compounds measured at 5 K and 300 K.

It is seen from Figure 6 (d) that the M-H curve of $Ti_{0.94}Mg_{0.06}O_2$ compound, which is measured under zero-field cooling (ZFC) condition, does not saturate even at 5 T field and shows linearity and vertical pinning kind behavior while reversing the applied magnetic field. It raised the possibility of manifestation of exchange bias due to the presence of domain wall pinning and due to the presence of competing ferromagnetic (FM) and the antiferromagnetic (AFM) interaction. In order to further understand these features, the M-H curves were measured at various temperatures as shown in the Figure 7. The variation of coercive field with temperature is presented in Figure 8 (a), where we can see that, $H_C$ value falls exponentially as the temperature is increased and it is a signature of dominant domains wall pinning [59]. Moreover, it is observed that the magnetic hysteresis loops exhibit a small shift along the magnetic field axes as shown in the inset of Figure 7 (a-h), implying the presence of an exchange bias effect in $Ti_{0.94}Mg_{0.06}O_2$ even in ZFC condition. The exchange bias field is calculated based on $H_E = (H_{C1} + H_{C2})$, where $H_{C1}$ and $H_{C2}$ are the coercive fields at the descending and the ascending branches of the magnetic hysteresis loop, respectively. Figure 8 (b) presents the variation of exchange bias field with temperature. The essential physics underlying to explain the exchange bias in core-shell structure is the exchange interaction between the antiferromagnet (core) and ferromagnet (shell) at their interface. Since antiferromagnets have a small or no net magnetization, their spin orientation is only weakly influenced by an externally applied magnetic field. FM shell which is exchange-coupled to the AFM (core) will have its interfacial spins pinned. Reversal of the ferromagnet's moment will have an added energetic cost corresponding to the energy necessary to create a domain wall within the antiferromagnetic. The added energy term implies a shift in the switching field of the ferromagnet. For a core/shell FM/AFM nanoparticle system, $H_E$ can be expressed as [60]:

$$H_E = 2 \frac{n J_{ex} S_{FM} S_{AFM}}{a^2 M_{FM} t_{FM}} \text{ --------------- (1)}$$

Here, $J_{ex}$ is the interfacial exchange constant, $S_{FM}$ and $S_{AFM}$ are individual spin moments of the FM shell and the AFM core respectively. $M_{FM}$ and $t_{FM}$ are the saturation magnetization and effective thickness of the ferromagnetic (shell) layer, and $n/a^2$ is the number of exchange-coupled bonds across the interface per unit area. Similar exchange bias behavior in other magnetic nanoparticles with a core–shell (or compound, or binary) structure have been studied extensively [57, 59-60].

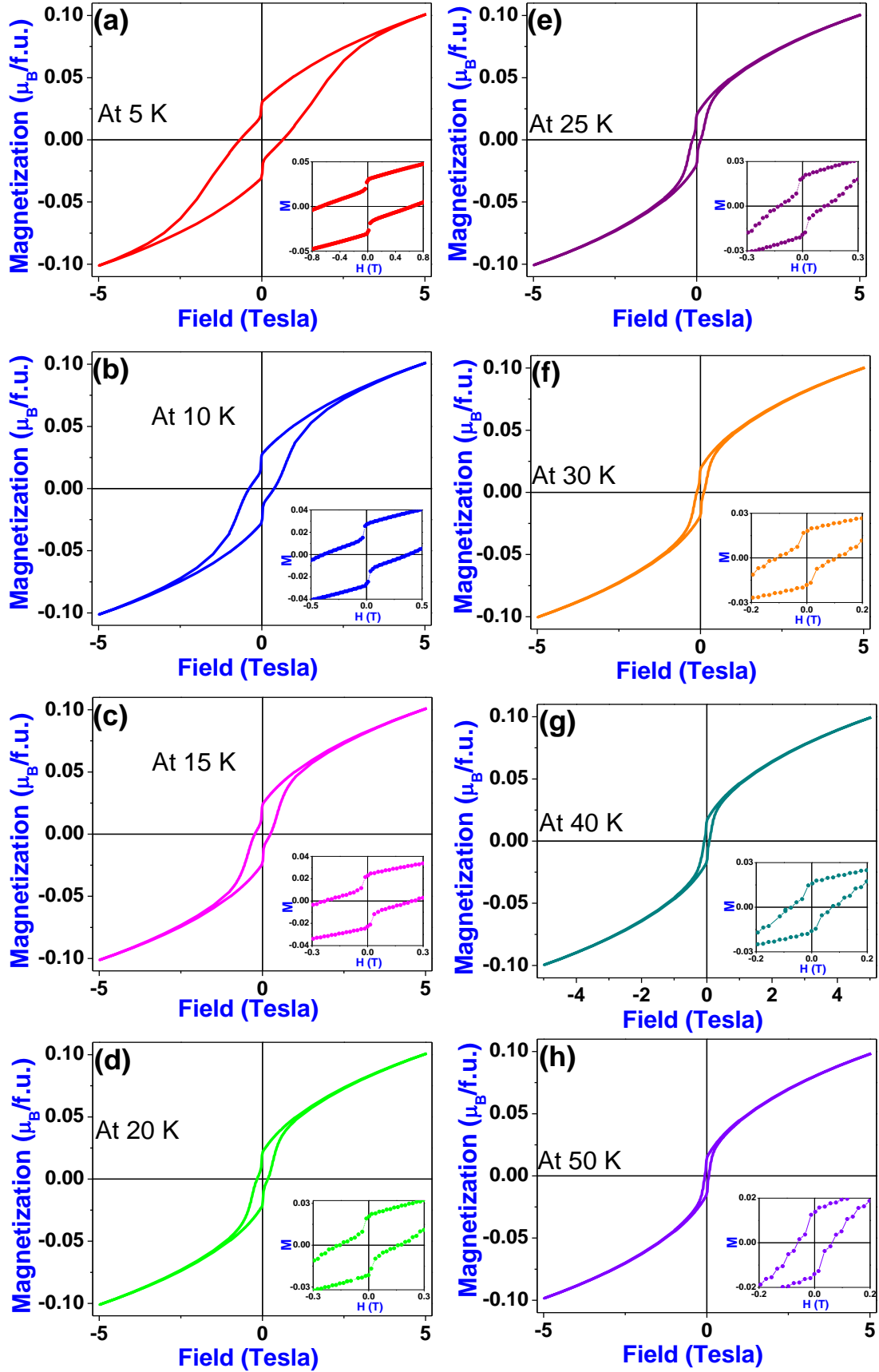

**Figure 7:** Magnetization versus field variation measured at various temperatures of $Ti_{0.94}Mg_{0.06}O_2$ compound.

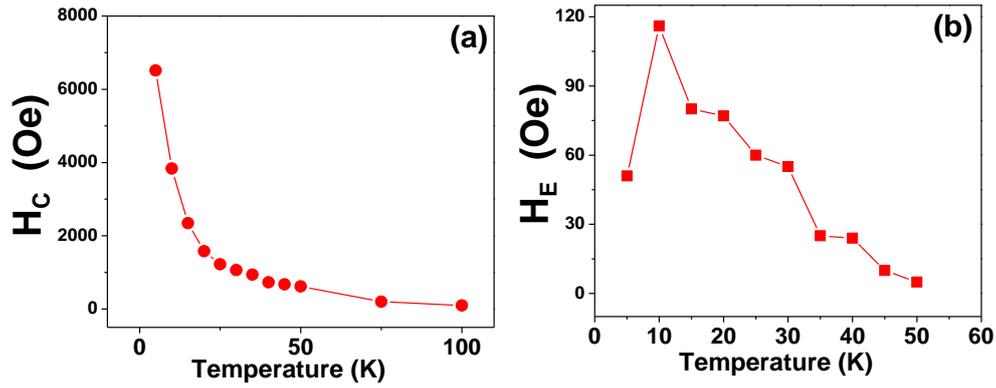

**Figure 8:** **(a)** Variation of coercivity **(b)** Exchange bias field with temperature of $Ti_{0.94}Mg_{0.06}O_2$ compound.

## 4. Conclusion

To conclude, the polycrystalline samples of $Ti_{1-x}Mg_xO_2$ (x=0, 0.02, 0.04 and 0.06) were prepared by solid-state reaction route. All the diffractions peaks could be indexed on the basis of the tetragonal rutile type-structure with single phase. The refinement of the XRD patterns with the help of the Fullprof program by employing the Rietveld refinement technique reveals no change in the crystallographic lattice parameters in comparison to pure $TiO_2$ upon Mg doping. This observation strongly indicates that no solid solution is formed i.e. that $Mg^{2+}$ ions do not enter grain cores and core/shell structure is formed. SEM observations reveal the uniform morphology with nanometric grains in the range of 150-200 nm. The measurement of magnetic properties of these compounds indicates that pure $TiO_2$ and $Ti_{0.98}Mg_{0.02}$ compounds exhibit paramagnetic behavior and $Ti_{0.96}Mg_{0.04}$ compound exhibits weak ferromagnetic phase superimposed with the dominating paramagnetic phase. However, $Ti_{0.94}Mg_{0.06}$ compound exhibits ferromagnetic to paramagnetic transition with FM transition temperature of 180.2 K. Moreover, the magnitude of magnetization is found to increase with Mg concentration. The measurements of ZFC and FC magnetization data for $Ti_{0.98}Mg_{0.02}O_2$ sample indicate low temperature magnetic irreversibility in ZFC and FC curves, which suggests two competing magnetic phases i.e. antiferromagnetic (AFM) phase with dominant ferromagnetic (FM) are present below irreversibility temperature. The observed magnetism was attributed to core (AFM) and the shell (FM) magnetism. The measurement of M-H curves of $Ti_{0.94}Mg_{0.06}O_2$ compound does not saturate even at 5 T field and shows linearity. The measurement of M-H data at various temperatures indicates domain wall pinning and exchange bias behavior.